\documentclass[sigconf]{acmart}

\copyrightyear{2024}
\acmYear{2024}
\setcopyright{rightsretained}
\acmConference[IDE '24]{2024 First IDE Workshop}{April 20, 2024}{Lisbon, Portugal}
\acmBooktitle{2024 First IDE Workshop (IDE '24), April 20, 2024, Lisbon, Portugal}\acmDOI{10.1145/3643796.3648452}
\acmISBN{979-8-4007-0580-9/24/04}

\begin{document}

\newcommand{\todo}[1]{\textcolor{red}{(TODO: #1)}}

\title{A New Generation of \emph{Intelligent} Development Environments}

\author{Mark Marron}
\affiliation{%
  \institution{University of Kentucky}
  \city{Lexington}
  \country{USA}
}
\email{marron@cs.uky.edu}

\begin{abstract}
    The practice of programming is undergoing a revolution with the introduction of AI assisted development (copilots) and the creation 
    of new programming languages that are designed explicitly for tooling, analysis, and automation. Integrated Development Environments (IDEs) 
    as they are currently conceptualized have not yet responded to these changes. They are still designed around the idea of a human programmer 
    typing textual code into an editor window with the IDE providing assistance via the integration of various tools for syntax highlighting, 
    compilation, debugging, and (maybe) code version control. This paper presents a vision for transforming the IDE from an \emph{Integrated} Development 
    Environment to an \emph{Intelligent} Development Environment. The new IDE will be designed around the idea of a human programmer as the manager or 
    curator of a software project who, rather than manually typing in code to implement a solution, will instead use the IDE to direct AI  
    programming agents and/or automated tools to combine existing APIs, packages, and new code to implement the needed features. In this new model, 
    the fundamental roles of the IDE are to 1) facilitate the communication between the human programmer and the AI agents and automated tools 
    and 2) organize the workflow tasks needed to go from requirements gathering to the final tested and validated deployed feature. This paper 
    presents a vision for the new Intelligent Development Environment based on a range of proof-of-concept high-value scenarios we have experimented with 
    and discusses the challenges that remain to realizing these in a cohesive intelligent development experience.

\end{abstract}

    

\maketitle

\begin{figure*}
    \centering
    \includegraphics[width=.8\textwidth]{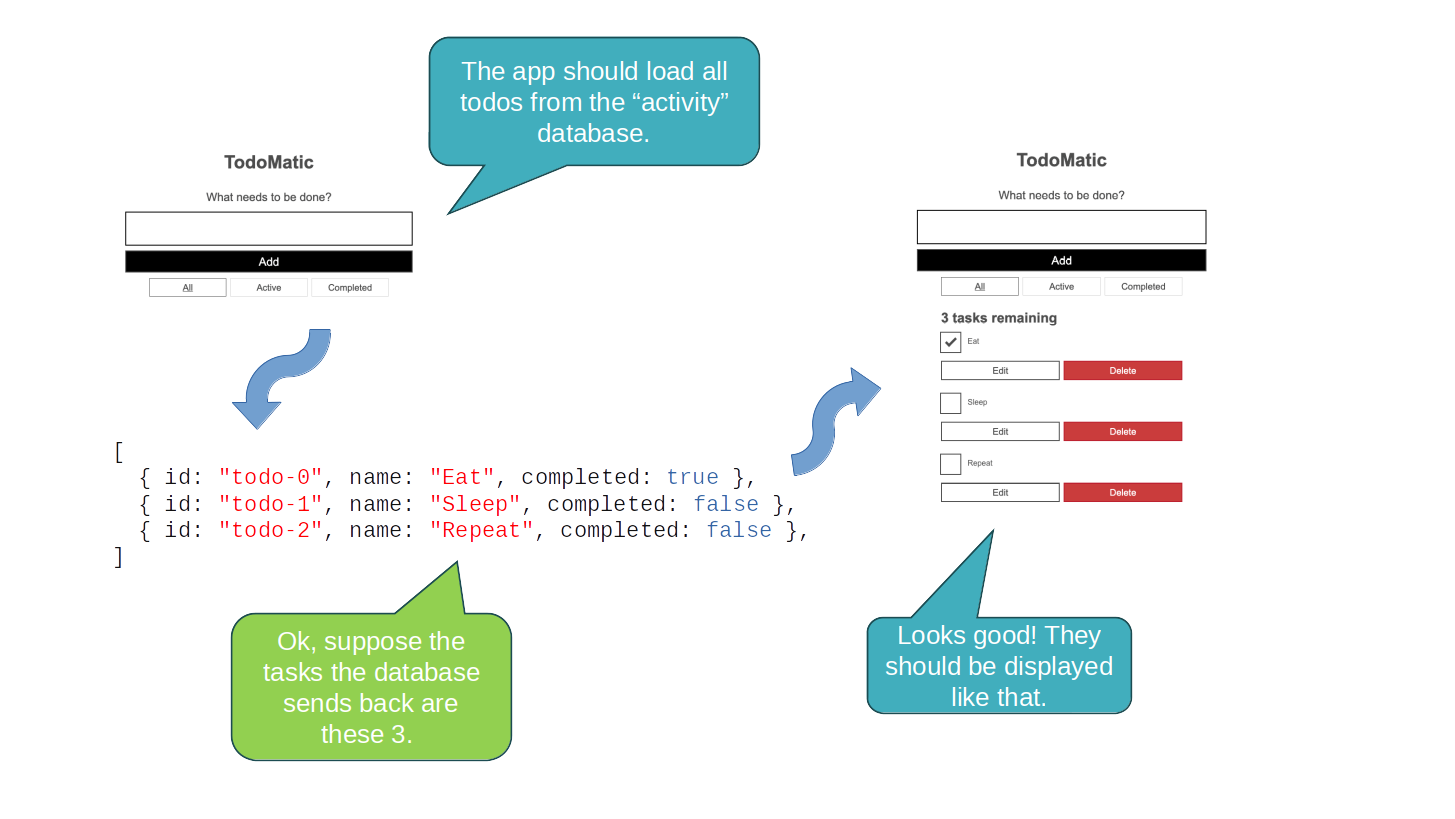}
    \caption{Requirements Gathering Workflow}
    \Description{Example of the live requirements gathering workflow and mock data.}
    \label{fig:requirements}
\end{figure*}

\section{Introduction}
The last 5 years have seen a revolution in the practice of programming. The introduction of AI assisted programming (copilots)~\cite{copilot} 
has changed the way that developers work by providing a new way to generate code. The rise of easy to use package ecosystems~\cite{npm} has 
made it easier than ever to reuse existing code and to integrate it into new applications. The introduction of new programming languages 
designed explicitly for tooling, analysis, and automation~\cite{bosque} has made the concept of transparent program validation and analysis 
a real possibility. The rise of distributed cloud programming, with serverless and REST based microservices, has has made it easier than ever to 
deploy and scale applications but created a whole new ecosystem of monitoring and observability tools. 

These changes have fundamentally altered the way that developers work. However, the Integrated Development Environments (IDEs) that developers 
use have not kept pace with these changes. They are still designed around the idea of a human programmer typing textual code into an editor 
window with the IDE providing some assistance via the integration of various tools for syntax highlighting, compilation, debugging, and (maybe) 
code version control. This paper presents experience from working on these technologies and a vision for transforming the IDE from 
an \emph{Integrated} Development Environment to an \emph{Intelligent} Development Environment.

This new IDE will be designed around the idea of a human programmer as the manager or curator of a software project who will use the development 
environment to gather requirements directly in a live and semantically rich way then to direct AI programming agents and/or automated tools to combine 
existing APIs, packages, and new code to implement the needed features. This environment also supports new workflows for validating and 
resolving potential issues in the code, before it is even integrated, ensuring that the final implementation satisfies key correctness, security, 
and performance requirements. Finally, in this new model, the development environment can fully close the lifecycle loop by working 
as an information broker between the developer and the deployed application. This allows the developer to quickly understand the current health 
of the application from a holistic perspective, to act on issues without requiring context switching between tools, and re-enter the requirements 
gather process seamlessly as new needs arise.

This paper presents a vision for the new \emph{Intelligent} Development Environment based on a range of proof-of-concept high-value scenarios 
we have experimented with and outlines the work needed to realize these features/scenarios in a cohesive development experience.

\section{Requirements Gathering}
\label{sec:requirements}
A key task for developers in the new world of AI assisted programming is to gather requirements and translate them into actionable 
tasks for the AI agents and automated tools. Our experiments with this task have identified a workflow based around Prorogued Programming~\cite{prorogued}, 
live UI mockups (using React~\cite{react}), and rich data description languages~\cite{typespec,bosqueapi}. 

In this workflow the developer interactively walks through scenarios with the customer to identify the overall workflows that should be 
supported using live UI mockups, written declaratively in React, and fills in the details of the data that is read/written in each step with 
mock values written in a rich data description language (TypeSpec~\cite{typespec}) as they are needed at each step.

\autoref{fig:requirements} shows an example of this workflow. In this scenario a developer is working to implement a new Todo application. 
The first action the customer wants on startup is to show the current todo tasks in the database. The developer 
can add the edge to a flow diagram, and add some free form text about the requirement + where to find the database and what login credentials 
should be used. The developer can then ask the customer what the data should look like and fill in the mock 
information, ideally with extensive autocompletion support if the format is available using techniques like~\cite{bosqueapi,typespec}. From this 
data a quick React UI component can be written and the customer can interactively walk through the flow to see if it matches their expectations. 
Further, as these flows are now captured in live-executable flow diagrams and UI components the developer and customer can interactively walk 
through them with different data scenarios and experiment with changes to the UI and/or planned computation components!

\section{AI Assisted Programming}
\label{sec:ai}
Once the requirements have been gathered and the data has been described the next step is to use AI agents to generate the code needed to 
realize the requirements. This is done by using the data description to generate a set of data-specifications that describe the APIs and 
data that each step operates on -- in our scenarios using TypeSpec~\cite{typespec} or Bosque~\cite{bosqueapi}. These specifications are 
then consumable by various AI agents, as a Mixture of Experts (MoE) that can generate code to implement the needed functionality. A first 
agent works to convert the flow diagram and data specifications into a set of high-level APIs that can be used to implement the functionality. 
Individual coding agents can then be used to implement each of the APIs, with the MoE agent providing guidance and feedback to the developer 
as they work. 

A key aspect of this workflow is that the developer is not writing code, but instead is managing the AI agents and providing feedback to 
refine and correct as the code is generated. This is a key shift in the role of the developer from a programmer to a manager/curator and 
this shift in roles must be supported in the development environment as well!

A first step in this direction is to support the automatic conversion of the sample data gathered as part of the requirements into unit 
tests that can reject obviously incorrect code. Another step is to use the API and data specifications to generate a set of sanity fuzz 
inputs to catch and correct any hard failures as the AI agents are working. However, the most important step is to provide direct support 
for the developer when they need to step in and provide feedback in situations where there is ambiguity in the requirements and the AI 
agents have identified multiple possible solutions. In these cases simply asking the developer to pick the correct solution by code inspection 
is not a viable option as there may be many possible solutions, with 100's of lines each, and small differences between versions! Instead,
the IDE can provide features~\cite{disambiguation} that show outputs which differ between the solutions, highlight key algorithmic 
differences in solutions, or that paraphrase key information back in natural language. These enable the developer to quickly identify the 
ambiguity and to provide feedback to the AI agents to resolve it.

\begin{figure}
    \centering
    \includegraphics[width=\columnwidth]{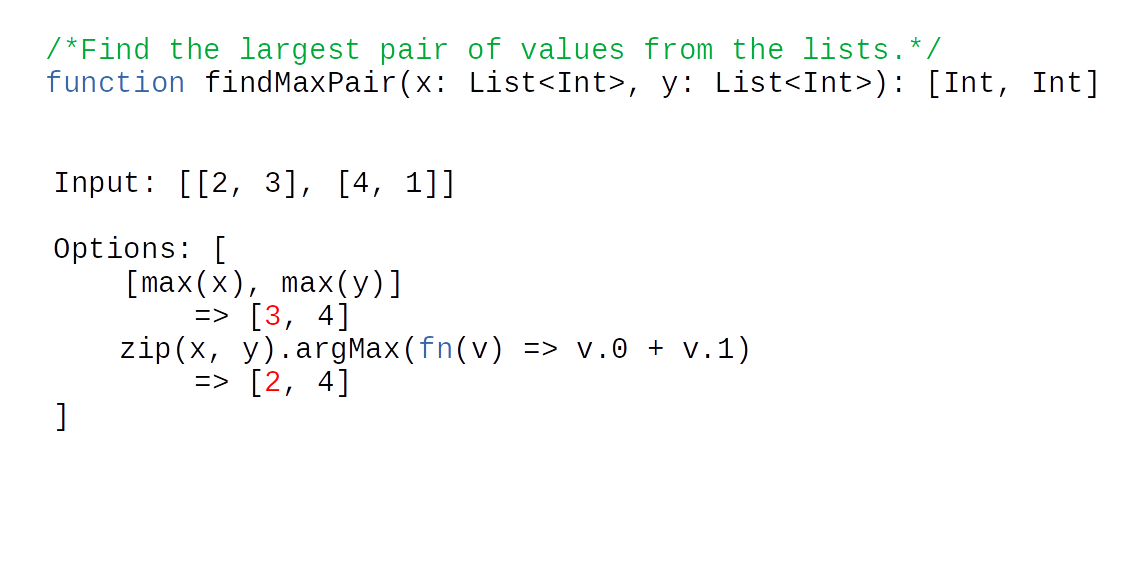}
    \caption{Disambiguation Support}
    \Description{Example of the disambiguation support.}
    \label{fig:disambiguation}
\end{figure}

An example of this is shown in \autoref{fig:disambiguation}. In this example the developer has asked the AI agents to generate a set of 
programs that implement the functionality described by the natural language ``find the max pair for the lists'', the signature given by 
the API description, and a single example/test case \texttt{[[1, 2], [2, 3]] => [2, 3]}. From this set of requirements an AI agent might 
generate two plausible solutions, one that is the max of the individual lists \texttt{[max(x), max(y)]} and that is the max sum of the 
index-wise pairs \texttt{zip(x, y).argMax(fn(v) => v.0 + v.1)}. 

Even for this simple code, manual inspection requires some effort to identify the differences between the two solutions and think about 
which one might be the intended one. However, the IDE can provide support to make this easier. In this case the IDE can show an input on 
which the two solutions differ and the output generated by each solution. In this case the developer, or even a non-expert customer, quickly 
identify which result is the desired one, and thus, which is the correct solution. Having IDE support in the form of overlays, side-by-side views 
with highlighting, or other UX affordances to highlight the key differences between solutions can make this process even easier.

\section{Deep Tooling Support}
\label{sec:deep}
A key aspect of the new Intelligent Development Environment is that it is designed to support the developer as a manager/curator of the 
AI agents and automated tools. In this role a developer needs to quickly understand the behavior of AI generated code or imported packages 
and validate that they are being used correctly. This requires a new generation of deep tooling support that can provide the developer 
with the information they need. Instead of simply providing a list of methods, types, and use/def sites that a developer must manually 
investigate to understand the requirements of an API call or what properties a type guarantees the IDE should be able to automatically 
generate this information -- and ideally automatically validate higher level issues as well including, if an API is used correctly or if 
there are possible runtime errors~\cite{bosque} or if SemVer~\cite{semver,svertalk} information is being updated consistently with a code change. 

\begin{figure}
    \centering
    \includegraphics[width=\columnwidth]{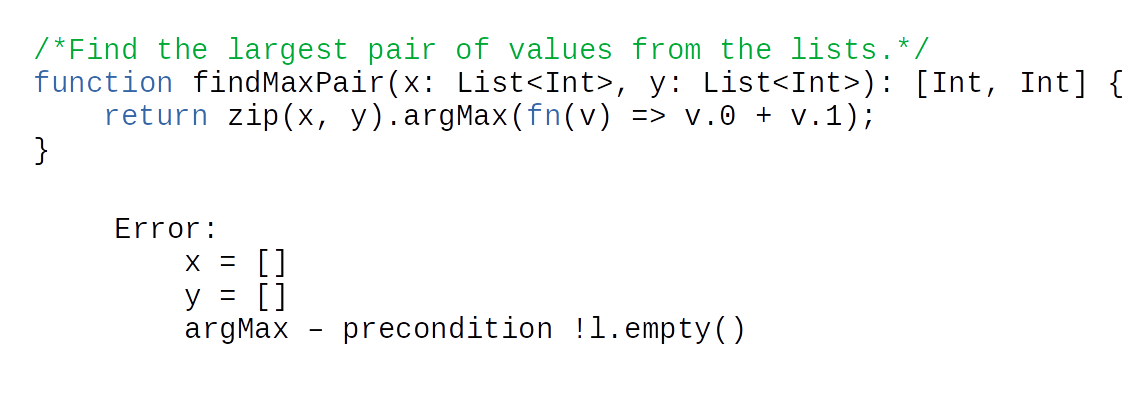}
    \caption{Example of automatically identifying possible runtime errors.}
    \Description{Example of tooling automaticaly identifying a possible runtime error.}
\label{fig:errorcheck}
\end{figure}

\autoref{fig:errorcheck} shows the running example code that has been generated by the AI agents, and after disambiguation from other options, 
selected by the developer. In this case the developer has asked the IDE to check if there are any possible runtime errors in the code. As 
the code is written in \textsc{Bosque}~\cite{bosque,fintech} the IDE can use a static validator to solve for possible failing inputs, or prove that 
none exist, for code like this. In this case there is a possible runtime error as the \texttt{argMax} function may be called with an empty list 
and the checker can automatically find this case for the developer!

However, it is critical that the development environment provide more than just red squiggles to indicate possible errors. Instead, we want to 
be able to provide the full input that causes the failure, key points in the code that are related, and ideally a direct way to jump into a 
debugger experience~\cite{ttt} to fully investigate the issue. In this case the IDE can show the input that causes the failure, the line of code that fails, 
and even suggest fixes to the spec to resolve the issue. 

\begin{figure}
    \centering
    \includegraphics[width=\columnwidth]{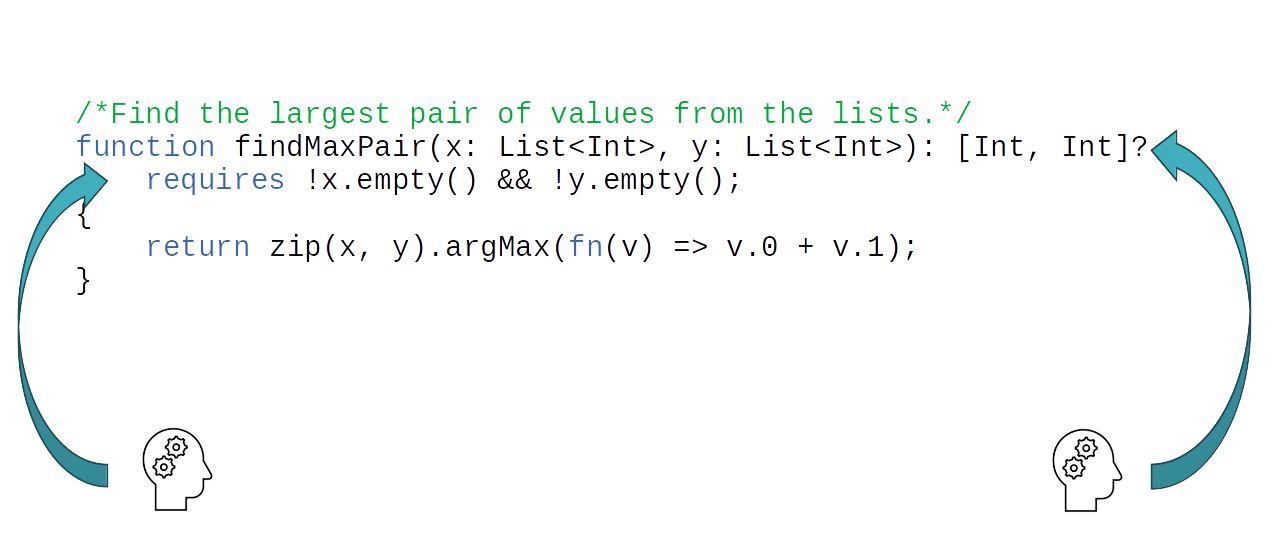}
    \caption{Example of automatically suggesting revisions to the specifications.}
    \Description{Example of automatically suggesting revisions to the specifications.}
\label{fig:fixopts}
\end{figure}

\autoref{fig:fixopts} shows the case where we can automatically identify and suggest refinements to the specifications to resolve the issue. In this 
example there are two options which the developer may want to consider, one adds a requirement that the inputs cannot be empty lists, 
\texttt{requires !x.empty() \&\& !y.empty()}, while the 
other relaxes the output type to include a \texttt{none} value for this special case. With support for the development environment to show 
these choices, and ideally the impacts they have on other parts of the codebase and semantic versioning information, the developer can quickly 
make an informed decision about which option to select.

While these examples have focused on correctness and runtime errors, the same approach can be used to provide support for other aspects of 
the development process. A key issue in many organizations is discoverability of APIs and packages. Using many of the same techniques, and 
richer specification languages~\cite{typespec,bosqueapi}, the IDE can provide support for semantically searching package databases for relevant 
APIs, types, and services. Being able to easily surface these into the development environment can make it easier for developers to reuse 
and leverage existing code and packages.

\section{Information Broker}
\label{sec:broker}
At this point in out hypothetical application (or feature) development process we can imagine that the developer has finally deployed the 
feature. Today, this is where the development process ends and the operations process begins. However, in the new world of an Intelligent 
Development Environment we envision that the IDE will be a one-stop-shop for all information about the application that is currently spread 
across logs, APM monitoring tools, SCM activities, and even telemetry and crash reports! 

This information allows a developer to quickly understand the current health of the application in a holistic way and act on issues without 
requiring context switching between tools. For example, if a developer is notified of a crash report they can quickly jump to the code line 
where the error is reported, see information on the number of times it has occurred, know about recent changes relevant code, and even jump 
directly into a debuggable recorded trace (if Time-Travel-Tracing is supported~\cite{ttt}) to investigate the issue.

Closing the loop back to the requirements gathering phase this integration provides a natural way to gather feedback from the customer and 
identify new requirements, to fix bugs, improve performance, or link open feature requests into the development environment. In our opening 
Todo example (\autoref{fig:requirements}) we can imagine that the customer has reported a bug where the application is not showing the correct 
data when there is a failure to connect to the database. The developer can quickly jump to the code that is responsible for this behavior, 
see the number of times it has occurred, triage the issue, and start modifying the requirements information to add error handling code. This 
then feeds into a new loop of AI assisted programming, validations, and then redeployment!

\balance

\section{Onward!}
\label{sec:onward}
Our experience with the scenarios and piece-wise implementations of the new Intelligent Development Environment has shown that there is 
great potential for this new model of development. However, there is substantial work to be done to realize this vision. In particular, 
many of the steps require:
\begin{itemize}
    \item \textbf{Extending the Environment Beyond Code} -- Many of the activities described in this work involve working with non-code, 
or non-traditional code, artifact such as natural language, flow diagrams, partial UI components, history of disambiguated code options, etc. 
This requires the IDE to be able to store and organize these artifacts in a first-class way -- not just treat them as random text files or blobs.
    \item \textbf{Programming Language Support} -- Many of the automation features described in this work require a new generation of programming 
languages, such as the Bosque work~\cite{bosque,bosqueapi} or TypeSpec~\cite{typespec}, that are designed to support tooling, analysis, and automation. As 
a result there is a feedback loop between development environment and programming language design that needs to be explored.
    \item \textbf{Rich Overlay or Specialized Modes} -- The new Intelligent Development Environment will need to support a range of new workflows. 
Our experience with various proof-of-concept workflows has shown that these workflows do not fit well into the traditional model of a tree of text 
files with a single editor window. Instead, we need to explore new ways to organize and present information to the developer. This may involve 
rich overlays, specialized modes, and even new UI paradigms.
\end{itemize}

In this future an Intelligent Development Environment will be more than just a combination of a text editor for a developer to type code into 
and some integration of various external, build, test, and debugging tools. Instead, it will be a new kind of development experience that 
is designed around the idea of a human programmer as the manager/curator of a software project who use the system to organize, explore, 
orchestrate, and validate the software development process.

These areas present a rich set of research challenges for the community to explore. However, we believe that the potential benefits of this 
work are substantial and will result in a new kind of development experience that is more productive, more accessible, and more enjoyable for 
developers and customers alike! 

\bibliography{bibfile}

\begin{thebibliography}{10}

\bibitem{prorogued}
Mehrdad Afshari, Earl~T. Barr, and Zhendong Su.
\newblock Liberating the programmer with prorogued programming.
\newblock In {\em Onward!}, 2012.

\bibitem{ttt}
Earl~T. Barr, Mark Marron, Ed~Maurer, Dan Moseley, and Gaurav Seth.
\newblock Time-{T}ravel {D}ebugging for {J}avascript/{N}ode.js.
\newblock In {\em FSE}, 2016.

\bibitem{bosqueapi}
Bosque object notation api.
\newblock \url{https://github.com/BosqueLanguage/BSQON}, 2023.

\bibitem{copilot}
Copilot.
\newblock \url{https://github.com/features/copilot}, 2023.

\bibitem{fintech}
Stephen Goldbaum, Attila Mihaly, Tosha Ellison, Earl~T. Barr, and Mark Marron.
\newblock High assurance software for financial regulation and business
  platforms.
\newblock In {\em VMCAI}, 2022.

\bibitem{semver}
Patrick Lam, Jens Dietrich, and David~J. Pearce.
\newblock Putting the semantics into semantic versioning.
\newblock In {\em Onward!}, 2020.

\bibitem{svertalk}
Mark Marron.
\newblock Agile and dependable service development with {B}osque and {M}orphir.
\newblock Linux Foundation Open Source Summit, 2021.

\bibitem{bosque}
Mark Marron.
\newblock Toward programming languages for reasoning: {H}umans, symbolic
  systems, and {AI} agents.
\newblock In {\em Onward!}, 2023.

\bibitem{disambiguation}
Mika\"{e}l Mayer, Gustavo Soares, Maxim Grechkin, Vu~Le, Mark Marron, Oleksandr
  Polozov, Rishabh Singh, Benjamin Zorn, and Sumit Gulwani.
\newblock User interaction models for disambiguation in programming by example.
\newblock In {\em UIST}, 2015.

\bibitem{npm}
Npm.
\newblock \url{https://www.npmjs.com/}, 2023.

\bibitem{react}
React.
\newblock \url{https://microsoft.github.io/typespec/}, 2023.

\bibitem{typespec}
Typespec.
\newblock \url{https://microsoft.github.io/typespec/}, 2023.

\end{thebibliography}
\bibliographystyle{plain}

\end{document}